\documentclass[a4paper, oneside, twocolumn, notitlepage, 10pt]{extarticle_ecoc}
\usepackage{ecoc}
\usepackage{float}
\usepackage{acronym}
\acrodef{EDFA}{Erbium-Doped Fiber Amplifier}
\acrodef{VOA}{Variable Optical Attenuator}
\acrodef{SELU}{Scaled Exponential Linear Unit}
\acrodef{NN}{Neural Network}
\acrodef{ReLU}{Rectified Linear Unit}
\acrodef{MSE}{Mean Squared Error}
\acrodef{MAE}{Mean Absolute Error}
\acrodef{TL}{Transfer Learning}
\acrodef{ML}{Machine Learning}
\acrodef{SNN}{Self Normalizing Neural Networks}
\begin{document}
\selectlanguage{english}

\title{Self-Normalizing Neural Network, Enabling One Shot Transfer Learning for Modeling EDFA Wavelength Dependent Gain}%
\vspace{-8mm}

\author{Agastya Raj\textsuperscript{*,}\textsuperscript{(1)},  
Zehao Wang\textsuperscript{(2)}, 
Frank Slyne\textsuperscript{(1)}, 
Tingjun Chen\textsuperscript{(2)}, 
Dan Kilper \textsuperscript{(1)}, 
Marco Ruffini\textsuperscript{(1)}}

\vspace{-6mm}

\maketitle

\begin{strip}

	\begin{author_descr}
		\textsuperscript{(1)} CONNECT Centre, School of Computer Science and Statistics and School of Engineering, Trinity College Dublin, Ireland,
		\textsuperscript{*}\textcolor{blue}{\uline{rajag@tcd.ie}} 
  
  \textsuperscript{(2)} Duke University, Department of Electrical and Computer Engineering, Durham, NC, USA
  \vspace{-3mm}
		
		
		
	\end{author_descr}
\end{strip}

\setstretch{1.063}
\renewcommand\footnotemark{}
\renewcommand\footnoterule{}


\begin{strip}
	\begin{ecoc_abstract}
We present a novel ML framework for modeling the wavelength-dependent gain of multiple EDFAs, based on semi-supervised, self-normalizing neural networks, enabling one-shot transfer learning. Our experiments on 22 EDFAs in Open Ireland and COSMOS testbeds show high-accuracy transfer-learning even when operated across different amplifier types. 	\textcopyright2023 The Author(s)	
\vspace{-3mm}
	\end{ecoc_abstract}
\end{strip}

\section{Introduction}
\vspace{-2mm}
The gain spectrum of an \ac{EDFA} has a complex dependence on channel loading, pump power, and operating mode, making accurate modeling difficult to achieve. Recently, \ac{ML} techniques such as \acp{NN} have been used to build \ac{EDFA} gain models~\cite{zhuMachineLearningBased2018, zhuHybridMachineLearning2020}. Other work~\cite{darosMachineLearningbasedEDFA2020} has produced generalized ML-based EDFA models using training datasets collected from multiple EDFAs of the same make and model, which are shown to achieve lower \ac{MAE} of the gain spectrum prediction across multiple devices of the same make. Although these models achieve high prediction accuracy, they do require a large number of measurements, which can be time-consuming and difficult to obtain if the EDFA is in a live network. Due to the complexity of the model, \ac{NN} also suffer from non-convex training criteria and local minima, which complicate the training process especially with limited number of measurements. 

\ac{TL} techniques\cite{zhuangComprehensiveSurveyTransfer2021a} have been recently used to try and mitigate this issue, by training a base model on one \ac{EDFA} and then using this to model different devices, by only using minimal additional data from the new device. Recently, it was demonstrated\cite{Wang:23} that a single EDFA model can be transferred between different EDFAs of the same type using only 0.5\% of the entire dataset, showcasing the potential for efficient model transfer in this domain. Yet, the application of transfer learning across amplifiers of different types (i.e., from a EDFA Booster base model towards an EDFA Preamp target model) requires further investigation. In addition, work to date has mostly relied on training data from external features, such as input power levels and output gain spectra, which may not fully capture the complex behaviour of \acp{EDFA}.

In this paper, we implement and study a novel semi-supervised, self-normalizing \ac{NN} approach (hereafter referred to as the SS-NN model) that characterizes the wavelength-dependent gain of an \ac{EDFA} using just 256 labeled measurements along with additional unlabeled data (which are easier to obtain). By incorporating internal \ac{EDFA} features that are typically available in commercial telecom equipment, our model can be transferred to different \ac{EDFA} types with only a single new measurement through transfer learning. We evaluate our approach on 22 different \acp{EDFA} across the Open Ireland (based in Dublin, Ireland) and PAWR COSMOS (based in Manhattan, USA) testbeds, achieving a \ac{MAE} within 0.14 dB for same-type transfers and 0.17 dB for cross-type transfers.
\begin{figure*}[t]
	\centering
	\includegraphics[width=1.0\linewidth]{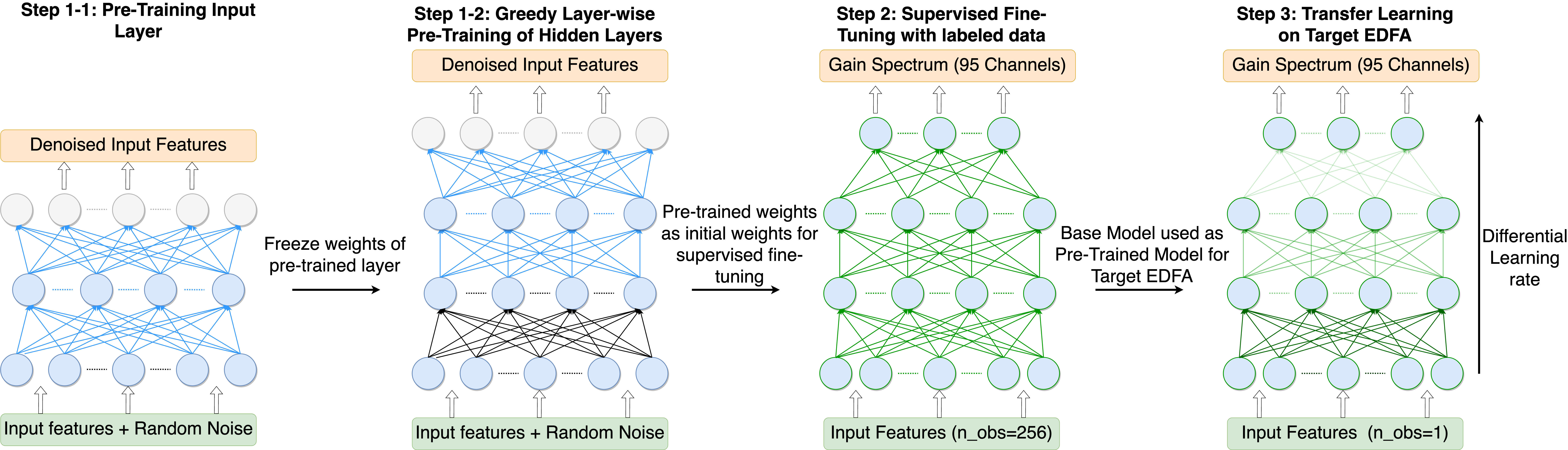}
	\caption{Model Training and Transfer Learning (TL) Framework}
	\label{fig:figure2}
\end{figure*}

\footnote{This paper is a preprint of a paper submitted to ECOC 2023 and is subject to Institution of Engineering and Technology Copyright. If accepted, the copy of record will be available at IET Digital Library}
\section{EDFA Gain Spectrum Measurement Dataset}

We carry out gain measurements across multiple wavelengths in the C-band from 3 commercial-grade Lumentum ROADM-20 units deployed in the Open Ireland testbed~\cite{open_ireland} and 8 similar units deployed in the PAWR COSMOS testbed~\cite{chenSoftwareDefinedProgrammableTestbed2022, raychaudhuri2020challenge}, each with 2 \acp{EDFA} (a Booster and a Pre-Amplifier), resulting in a total number of 22 \acp{EDFA}. To ensure consistency, we followed a similar measurement setup and data collection pipeline for both testbeds~\cite{wangOpenEDFAGain2023}. 
In the Open Ireland testbed, all \acp{EDFA} were measured at target gains of 15/20/25 dB, while in the COSMOS testbed, the target gains were 15/18/21 dB for Boosters and 15/18/21/24/27 dB for Pre-Amplifiers in high gain mode with 0 dB gain tilt (we adopt different gain setting to emulate diversity of operation in different networks). The dataset includes 3,168 gain measurements (at multiple wavelengths) for each \ac{EDFA}, for each given target gain settings, across 95$\times$50 GHz channels in the C-band. 
In addition, measurements for each \acp{EDFA} are collected under two channel loading modes: Random and Goalpost allocation (i.e., loading groups of channels in different spectrum bands).

\section{Training of the Base Model}

We construct a base model (used as the \emph{source model} in the transfer learning process) for both booster and pre-amplifier \acp{EDFA} using a 5-layer \ac{NN} with neurons initialized with zero weights. The \ac{NN} architecture consists of 200 neurons in the first two layers, 100 neurons in the next two layers, and 95 neurons in the final layer, predicting the wavelength-dependent gain output. Input features to the model include \ac{EDFA} target gain setting, total input/output power, input power for each channel, a binary vector for each channel representing the channel loading configuration, and three internal features related to the embedded \ac{VOA}: total \ac{VOA} input/output power and attenuation.

Due to the limitations of batch normalization when fine-tuning models with less than 32 observations~\cite{NIPS2017_c54e7837}, we utilize \ac{SNN} with \ac{SELU} activation function~\cite{klambauerSelfNormalizingNeuralNetworks2017a} instead of the \ac{ReLU}. This choice enables us to effectively normalize the hidden layer outputs with a small amount of data, while maintaining the benefits of hidden layer normalization and preserving high accuracy. This step is the key enabling factor of our developed \ac{NN} architecture to achieve effective one-shot training and transferability between models. 

We employ a 2-step process to train the source model, including unsupervised pre-training~\cite{NIPS2006_5da713a6}\cite{geProvableAdvantageUnsupervised2023} and supervised fine-tuning~\cite{s22114157} (see Fig.~\ref{fig:figure2}):

First, in the \emph{unsupervised pre-training} step, we initialize the source model's weights using unlabeled data from 512 measurements for each target \ac{EDFA} gain setting. Noise is added to the data, and the input layer is trained as an autoencoder to denoise and reconstruct the input. We construct this autoencoder by removing subsequent layers and adding a decoder layer on top. The autoencoder is trained for 1,800 epochs with a learning rate of 0.001, using the Adam Optimizer and \ac{MSE} loss function. The weights of this layer are fixed and used as the basis for training the subsequent layers.

Second, in the \emph{supervised fine-tuning} step, we utilize 256 randomly loaded gain spectrum measurements to train the model in a supervised manner. The model is trained using the \ac{MSE} loss function across all loaded channels, with a learning rate of 0.001 over 1,200 epochs. The test set comprises all Fixed Goalpost (270 measurements) and 20\% of the Random Baseline (220 measurements) \ac{EDFA} gain spectrum measurements, so that we can compare the performance for the two different channel loading scenarios\cite{Wang:23}.

\begin{figure*}[t]
	\centering
	\includegraphics[width=1.0\linewidth]{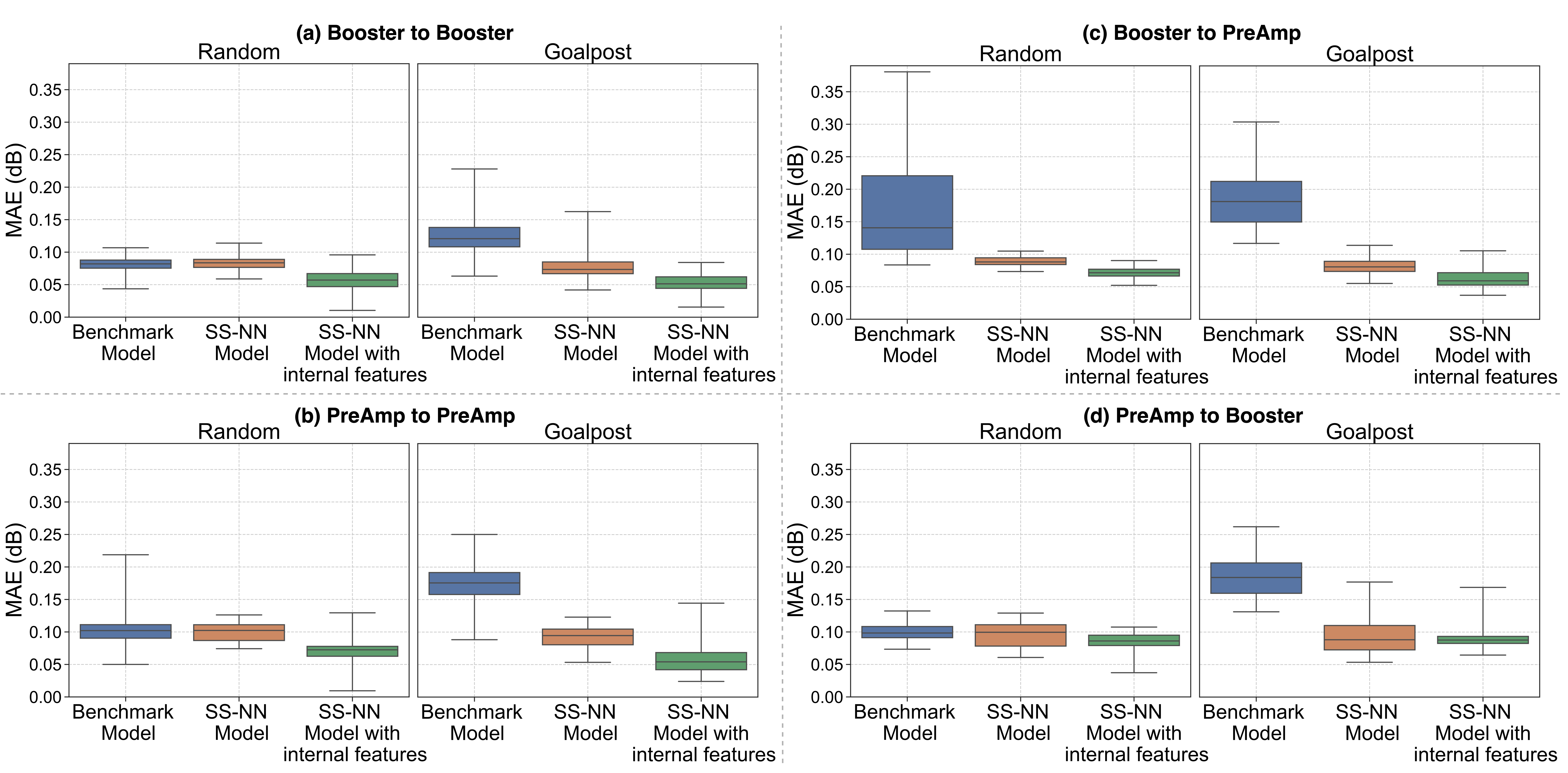}
	\caption{Boxplot distribution of \ac{MAE} across 22 \acp{EDFA} of (a) Booster to Booster \ac{TL}, (b) PreAmp to Preamp \ac{TL}, (c) Booster to Preamp \ac{TL} and (d) Preamp to Booster \ac{TL}. The boxes denote the inter-quartile range, and the whiskers denote the min/max.}
	\label{fig:figure3}
\end{figure*} 
\begin{figure*}[t]
	\centering
	\includegraphics[width=1.0\linewidth]{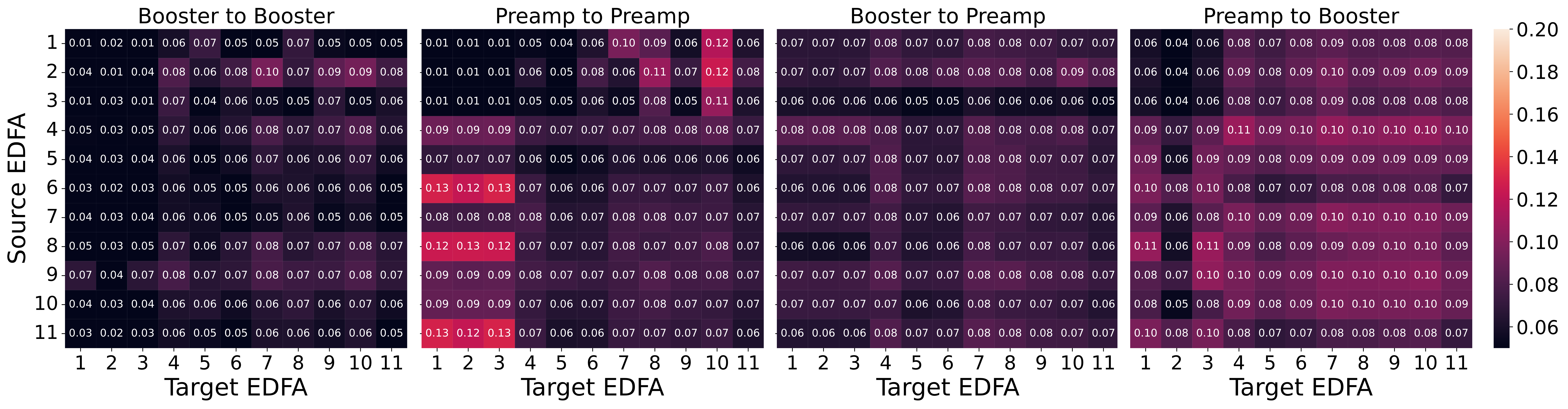}
	\caption{Transfer Learning \ac{MAE} matrix of SS-NN model with internal features on random loading. The (i, j) entry corresponds to the \ac{TL}-based \ac{EDFA} model, where the i\textsuperscript{th} and j\textsuperscript{th} \ac{EDFA} serve as the source and target models, respectively. EDFA 1-3 are deployed in Open Ireland, while EDFA 4-11 are deployed in COSMOS.}
	\label{fig:figure4}
\end{figure*}

\section{Transfer Learning (TL) to Target EDFA}

To transfer from a source model to a target \ac{EDFA}, we re-train the same model using a single randomly loaded measurement for 10,000 epochs, using \ac{MSE} as the loss function and Adam optimizer. We use a differential learning rate across layers, where the output layer has a larger learning rate of 1e-03 compared to the subsequent hidden layers, which have progressively decreasing learning rates, with each layer's rate being 10\% of the next layer's rate. In this way, the weights of the output layer are modified more aggressively, allowing it to capture the specific characteristics of the target \ac{EDFA} more effectively. At the same time, the lower levels of the \ac{NN} are fine-tuned more gradually to avoid overfitting and ensure that the model can be generalized to new inputs.

\section{Results}

We compare our SS-NN based \ac{TL} technique with a benchmark state-of-the-art method~\cite{Wang:23}, using the same set of features to highlight the benefit of our approach. Additionally, we demonstrate the advantage of incorporating internal \ac{EDFA} features by comparing the results of the SS-NN model with and without these additional features. 

\noindent\textbf{\ac{TL} to the Same \ac{EDFA} Type}:
Figs.~\ref{fig:figure3}(a) and~\ref{fig:figure3}(b) present the \ac{MAE} of the three approaches for \ac{TL} for source Booster{\textrightarrow}Target Booster and Source Pre-Amplifier{\textrightarrow}Target Pre-Amplifier respectively, for both random and goalpost channel loading, across 22 \acp{EDFA}. Our SS-NN model outperforms the benchmark technique for goalpost loading and exhibits comparable performance for random channel loading. In addition, when we include additional features, we see improvement in both channel loading configurations. 

\noindent\textbf{\ac{TL} to Cross-\ac{EDFA} Types}:
Figs.~\ref{fig:figure3}(c) and~\ref{fig:figure3}(d) report the \ac{MAE} of the three approaches for \ac{TL} between different types of \acp{EDFA} (Booster{\textrightarrow}Preamp and Preamp{\textrightarrow}Booster) respectively, under random and goalpost channel loading configurations. The SS-NN model again displays significant improvement and consistency in cross-type transfer, with a 3$\times$ improvement in \ac{MAE} over the benchmark algorithm even if using the same features. Including internal \ac{VOA} features further improves performance, leading to similar performance as the same-type transfer. 

A key advantage of the SS-NN model is its consistent performance across \ac{TL} between all \acp{EDFA}. Fig.~\ref{fig:figure4} shows the \ac{MAE} of the \ac{TL} model incorporating internal features on random channel loading. The model demonstrates consistent performance in terms of the \ac{EDFA} gain spectrum prediction accuracy, with an \ac{MAE} within 0.13 dB for same-type transfers and within 0.11 dB for cross-type transfers. 
\section{Conclusions}
\vspace{-2mm}

We analyze a novel semi-supervised learning technique to model the gain spectrum of an \ac{EDFA} using a minimal amount of data. The model can be transferred to \acp{EDFA} of different types using a single new measurement, showing that a single \ac{EDFA} can be used to characterize multiple \acp{EDFA} using minimal data collection. We also find that using internal \ac{EDFA} features available to the operator provides enhanced performance in both same-type and cross-type transfers, showing potential for improvement by incorporating internal features.

\section{Acknowledgements}

Supported by grants from SFI: 12/RC/2276\_p2, 18/RI/5721, and 13/RC/2077\_p2 and NSF: CNS-1827923, OAC-2029295, and CNS-2112562.

\printbibliography

\vspace{-4mm}

\end{document}